\documentclass [conference]{IEEEtran}
\IEEEoverridecommandlockouts
\usepackage{balance}
\usepackage{cite}
 \usepackage[keeplastbox]{flushend}
 \usepackage[utf8]{inputenc}
     \usepackage{setspace}
\usepackage{amsmath,amssymb,amsfonts}
\usepackage[compact]{titlesec}         
\usepackage{algorithmic}
\usepackage{graphicx}
\usepackage{subcaption}
\usepackage{epstopdf} 
\usepackage{textcomp}
\usepackage{xcolor}
\newcommand{\RN}[1]{%
\textup{\uppercase\expandafter{\romannumeral#1}}
}
\def\BibTeX{{\rm B\kern-.05em{\sc i\kern-.025em b}\kern-.08em
    T\kern-.1667em\lower.7ex\hbox{E}\kern-.125emX}}
\begin{document}
\title{On the Use of HAPS to Increase Secrecy Performance in Satellite Networks}
\IEEEoverridecommandlockouts  
\author{\IEEEauthorblockN{Olfa Ben Yahia\IEEEauthorrefmark{1}, Eylem Erdogan\IEEEauthorrefmark{2}, G{\"{u}}ne{\c{s}}~Karabulut~Kurt\IEEEauthorrefmark{1}\IEEEauthorrefmark{3}}

\IEEEauthorblockA{\IEEEauthorrefmark{1}Department of Electronics and Communication Engineering, Istanbul Technical University, Istanbul, Turkey}

\IEEEauthorblockA{\IEEEauthorrefmark{2}Electrical and Electronics Engineering, Istanbul Medeniyet University, Istanbul, Turkey} 
 \IEEEauthorblockA{\IEEEauthorrefmark{3}Department of Electrical Engineering, Polytechnique Montréal, Montréal, QC, Canada}\\
\textit{\{yahiao17, gkurt\}@itu.edu.tr,} \textit{eylem.erdogan@medeniyet.edu.tr,} \textit{gunes.kurt@polymtl.ca }}

\maketitle
\addtolength{\topmargin}{0.2in}
\begin{abstract}
In this paper, we investigate the secrecy performance of radio frequency (RF) eavesdropping for a high altitude platform station (HAPS) aided satellite communication (SatCom) system. More precisely, we propose a new SatCom scheme where a HAPS node is used as an intermediate relay to transmit the satellite's signal to the ground station (GS). In this network, free-space optical (FSO) communication is adopted between HAPS and satellite, whereas RF communication is used between HAPS and GS as the line-of-sight (LoS) communication cannot be established. To quantify the overall secrecy performance of the proposed scheme, closed-form secrecy outage probability (SOP) and the probability of positive secrecy capacity (PPSC) expressions are derived. Moreover, we investigate the effect of pointing error and shadowing severity parameters. Finally, design guidelines that can be useful in the design of practical SatCom networks are presented.
\end{abstract}


\section{Introduction}
Satellite communication (SatCom) is considered to provide high power and high frequency from everywhere at any time, independently from geography and infrastructure. It has been extensively used in various fields including broadcasting, disaster relief, and navigation due to its capacity of offering services over a wide area. Satellites are categorized into 3 groups depending on their distance to Earth; geostationary Earth orbit (GEO), medium Earth orbit (MEO), and the low Earth orbit (LEO) satellites. High orbit satellites are characterized by a weak resolution and higher latency, which makes them inadequate for delay-sensitive traffic. To satisfy the high demand for new technologies and avoid large delays, the use of LEO satellites has attracted much interest from academia. 
Traditionally, SatCom is based on radio-frequency (RF) systems, where different frequencies are used based on the application type. However, RF systems are prone to many issues including spectrum congestion, licensing problems, and interference with other bands. Moreover, RF SatCom is more susceptible to security risks due to the possibility of jamming and interception, which are critical especially for military communications. To eradicate these problems, researchers have focused on exploiting free-space optical (FSO) communications.
In fact, FSO SatCom can provide relevant advantages compared to its RF counterpart thanks to its unique characteristics. Unlike RF, FSO is considered to offer extremely high bandwidth, unlicensed spectrum, improved link security, and easy deployment \cite{li2019physical}.
In FSO SatCom, the main necessity is a clear line of sight (LoS) connectivity and accurate alignment between communicating parts to guarantee perfect communication. Additionally, it has been proven that the quality of laser SatCom can be affected by different factors including climatic conditions (heavy rain, clouds, and fog), operating frequency, and poor angle of inclination \cite{li2019physical}, which decrease the visibility range and cause attenuation in the propagated signal. The principal effects of turbulence on FSO downlink laser signals in SatCom are the scintillations and angle of arrival fluctuations. Furthermore, the direct communication between the satellite and the terrestrial users may be inaccessible due to the masking effect resulting from barriers or shadowing between them. To overcome these problems, RF communication can be used with its FSO counterpart in the so-called mixed RF-FSO relaying mode \cite{samimi2013end}. In this mode, FSO and RF communication can be employed together in a dual-hop configuration, where a high altitude platform station (HAPS) can be used as an intermediate relay node with the aid of the well-known amplify and forward (AF) or decode and forward (DF) relaying techniques.

The use of the HAPS systems in the aid of SatCom has attracted much interest both from academia and industry. A HAPS system can be defined as a network node operating at the stratosphere layer, which can provide significant advantages to the laser SatCom including high coverage, better signal quality, high reliability, higher throughput, and lower latency due to its small footprint \cite{kurt2020vision}. Moreover, stratosphere layer is considered to be less affected by harsh weather conditions and more secure for applications. On the contrary, the zenith angle and wind speed can be critical in the design of HAPS systems.

In the next generation of non-terrestrial networks, one of the most important concerns arises in the physical layer security. Specifically, any legitimate user which is positioned close to the ground station (GS) can capture the RF signal in the FSO-RF SatCom systems. Thus, physical layer security has been drawing much interest from academia and industry, as it provides perfect secrecy based on channel characteristics and imperfections. Several solutions have been proposed to secure the SatCom in [4]-[8]. \nocite{vazquez2017information} \nocite{bankey2018physical}\nocite{lei2018secrecy}\nocite{liu2018novel}\nocite{bankeyphysical}In the literature, Guo \textit{et al}. \cite{guo2016secure} studied a wiretap satellite-terrestrial network for shadowed-Rician channel and proved that under heavy fading, the system implies worse security performance. In addition, secrecy capacity was investigated for satellite links under the additive white Gaussian noise (AWGN) channel and rain attenuation in \cite{petraki2010secrecy}. Furthermore, \cite{vazquez2019one} and \cite{hayashi2020physical} proposed two-way protocols that guarantee secure communication compared to the normal wiretap channel without any assumption on the eavesdroppers' channel. Lei \textit{et al}. in \cite{lei2011secure} elaborated a novel system based on joint power control and beamforming that improves physical layer security and minimizes the satellite transmit power. Two beamforming systems and interference are implemented to improve the secrecy capacity and decrease the total transmit power in the cognitive satellite network in \cite{an2016secure} and \cite{lin2018joint}. Moreover, the secrecy outage probability (SOP) of a multi-antenna satellite-terrestrial system is studied for random positions of the GS and eavesdropper \cite{zhang2020secrecy}.
 
As the aforementioned studies show, using an intermediate node is a prevalent technique to enhance security performance and to extend the coverage area of SatCom. In \cite{ai2019physical}, the authors studied the average secrecy capacity and the SOP as performance metrics for hybrid satellite and the free-space-cooperative system in the cases of AF and DF relaying. Besides, to minimize the SOP, power allocation is optimized based on relay selection method in \cite{la2019physical}, whereas in \cite{miridakis2014dual}, dual-hop communication was investigated where the legitimate users are terrestrial nodes that communicate by the mean of a satellite relay over shadowed-Rician channels.

Motivated by the opportunities of FSO SatCom, we consider a HAPS-aided satellite network and derive overall secrecy performance. The main contributions of the paper are summarized as follows:
\begin{itemize}
\item Different from the previous works that consider RF-FSO hybrid systems, we consider a HAPS-aided FSO-RF SatCom, where the illegitimate user is located on the ground level.
\item We evaluate the security performance of HAPS-aided downlink FSO-RF SatCom. Furthermore, we emphasize the use of the HAPS to mitigate the effect of weather conditions on the FSO link.
\item We elaborate on the effect of the severity of fading, the severity of scintillation, and the zenith angle. Furthermore, we also consider the impact of pointing errors and heterodyne detection (HD) technique in the proposed scheme.

\item We derive the closed-form of SOP and the probability of positive secrecy capacity (PPSC) expressions by considering RF eavesdropping and provide design guidelines that can be helpful for the design of downlink SatCom.
\end{itemize}

The rest of this paper is organized as follows: In Section $\RN{2}$, the investigated system model and corresponding channel models are briefly introduced. The SOP and PPSC analyses are provided in Section $\RN{3}$. The analytical expressions are
evaluated and discussed in Section $\RN{4}$ along with the proposed design guidelines. Finally, Section $\RN{5}$ concludes the paper.

\vspace{0.3cm}
\section{Signals and system model}
\begin{figure}[!t]
  \centering
    \includegraphics[width=3in]{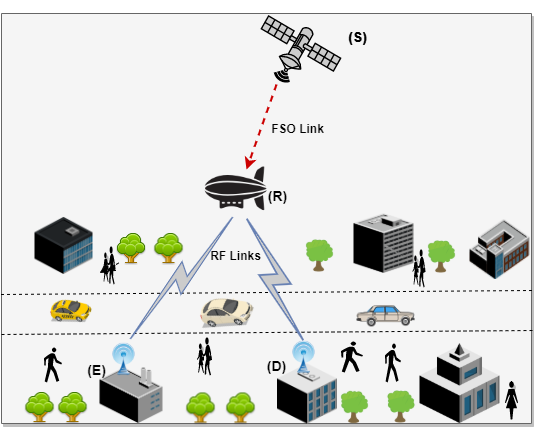}
  \caption{Illustration of the HAPS-aided FSO-RF SatCom system model.}
  \label{fig:model}
\end{figure}
In this paper, we investigate a HAPS-aided FSO-RF SatCom system as shown in Fig. 1. In this scenario, a LEO satellite $S$ seeks to communicate with a terrestrial destination $D$ through a HAPS node $R$ in the presence of a terrestrial passive eavesdropper $E$ that attempts to detect the confidential message sent by $S$. The communication link between $S$ and $R$ is through an FSO link that follows the Gamma-Gamma distribution, whereas the RF links are modeled as shadowed-Rician fading. In the first phase of the communication, the received signal can be expressed as 
\begin{align}
y_z=h_z x+n , z\in \left\lbrace RD,RE\right\rbrace,
\end{align}
where $x$ denotes the transmit signal vector and, $n$ is the AWGN, and $h_z$ is the channel response, which can be expressed as \cite{guo2018secrecy}
\begin{align}
h_z=F_z g_z,
\end{align}
where $g_z$ denotes the channel coefficient and $F_z$ is the scaling parameter including free space loss (FSL) and antenna pattern \cite{guo2018secrecy}. Considering DF relaying scheme, the overall signal-to-noise ratio (SNR) of our system can be expressed as 
\begin{align}
\gamma_0 = 
\min(\gamma_{SR}^{FSO},\gamma_{RD}^{RF}) , 
\end{align}
where $\gamma_{SR}^{FSO}=\frac{P_s}{N_0} I_{SR}^2$, $\gamma_{RD}^{RF}=\frac{P_R}{N_0}|h_{RD} |^2$.
\subsection{Channel Models}
Since the RF links experience the shadowed-Rician fading, the probability density function (PDF) and the cumulative distribution function (CDF) of the received SNR are given by \cite{ai2019physical}
\small
\begin{align}
&f_{\gamma_z}^{RF} (\gamma)= \sum_{k=0}^{m_z -1} \frac{\alpha_z(1-m_z)_k.(-\delta_z)^k \gamma^k}{\overline{\gamma}_z^{k+1} .(k!)^2} \exp(-\lambda_z \gamma), \\ \nonumber \\
&F_{\gamma_z}^{RF} (\gamma)= 1- \sum_{k=0}^{m_z -1} \sum_{i=0}^{k} \frac{\alpha_z(1-m_z)_k.(-\delta_z)^k \gamma^i}{i! \lambda_z^{k-i+1}\overline{\gamma}_z^{k+1} .k!} \exp(-\lambda_z \gamma),
\end{align}
\normalsize
where $\lambda_z=\frac{\beta_z - \delta_z}{\overline{\gamma}_z}$, $\overline{\gamma}_z$ is the average SNR for the channel $h_z$. $\alpha_z=\frac{1}{2b_z}(\frac{2 b_z m_z}{2 b_z m_z + \Omega_z})^{m_z}$, $ \beta_z=\frac{1}{2b_z} $, $\delta_z=\frac{\Omega_z}{2b_z(2b_z m_z + \Omega_z}$ with $m_z$ representing the Nakagami parameter of the corresponding link, and $\Omega_z$, $ 2b_z $ are the average power of the LoS component and multipath component. Finally, $(\cdot)_k$ denotes the Pochhammer symbol.

For the FSO link, we consider a clear alignment between the $S$ and $R$ apertures. Thereby, the PDF and CDF of the instantaneous SNR at the $R$ node ($\gamma_{SR}^{FSO}$) can be given as \cite{ai2019physical}
\small
\begin{align}
  \begin{array}{c}
f_{\gamma_{SR}}^{FSO} (\gamma)= \frac{\xi^2}{r \Gamma(\alpha)\Gamma(\beta) \gamma}
 G_{1,3}^{3,0}\left( h\alpha\beta(\frac{\gamma}{\mu_r})^{\frac{1}{r}} \middle \vert \begin{array}{c} 
\xi^{2+1} \\
\xi^2,\alpha,\beta 
\end{array}
\right),
\end{array}
\end{align}
\small
\begin{align}
  \begin{array}{c}
F_{\gamma_{SR}}^{FSO} (\gamma)= \frac{\xi^2 r^{\alpha+\beta-2}}{2\pi^{r-1} \Gamma(\alpha)\Gamma(\beta)}  
 G_{r+1,3r+1}^{3r,1}\left(\frac{(h\alpha\beta)^r} {\mu_r r^{2r} }\gamma \middle \vert \begin{array}{c}
1,k_1 \\
k_2,0 \end{array}
\right),
\end{array}
\end{align}
\normalsize
where the parameter $r$ indicates the used type of detection at the HAPS node (i.e., $r= 1$ for HD and $r= 2$ denotes intensity modulation with direct detection (IM/DD)), the coefficient $\xi$ implies the pointing error displacement at the receiver, $\alpha$ and $\beta$ specify the severity of fading and scintillation produced by the atmospheric turbulence conditions, $h=\frac{\xi^2}{\xi^{2+1}}$, $\mu_r$ presents the average SNR of the corresponding FSO link within the specific type of detection, and
$G_{p,q}^{m,n} \left( x \middle \vert \begin{array}{c}
a_1,…a_p \\
 b_1,…b_q \end{array}\right)$ 
is the Meijer G-function \cite{Wolform}. Furthermore, $k_1=\Delta(r,\xi^2+1)$, $k_2=\Delta(r,\xi^2),\Delta(r,\alpha),\Delta(r,\beta)$ where
the notation $\Delta(k,a)$ denotes $\Delta(k,a)= \frac{a}{k},\frac{a+1}{k}...,\frac{a+k-1}{k}$ including $k$ values. For simplicity, we assume the following notations hereinafter: $B= \frac{\xi^2 r^{\alpha+\beta-2}}{2\pi^{r-1} \Gamma(\alpha)\Gamma(\beta)}$, $D= \frac{(h\alpha\beta)^r} {\mu_r r^{2r}}$.
According to [\nocite{andrews2005laser}22, Sect. (12)], the severity of fading and the severity of scintillation parameters are defined by
 \begin{align}
  \alpha=
\left\lbrace  \exp \Bigg[ \frac{0.49 \sigma_R^2}{(1+1.11 \sigma_R^ {12/5}) ^{7/6}} \Bigg] - 1  \right\rbrace  ^{-1},  
\end{align}
  \begin{align}
  \beta=\left\lbrace \exp \Bigg[\frac{0.51 \sigma_R^2}{(1+0.69 \sigma_R^{12/5})^{5/6}} \Bigg] -1 \right\rbrace   ^{-1} ,
\label{EQN:9}
\end{align}
where $\sigma_R$ is the Rytov variance given as [22, Sect. (12)] 
\small
 \begin{align}
     \sigma_R^2=2.25 K^{7/6} sec^{11/6}(\zeta) \int_{h_0}^{H} C_n^2(h)(h-h_0)^{5/6} dh,
   \end{align}
 \normalsize
where $K=\frac{2\pi}{\lambda}$ is the wave number, $\lambda$ is the wavelength, $\zeta$ represents the zenith angle, and $C_n^2(h)$ is the refractive-index structure parameter expressed as \cite{ITUR}
      \begin{align}
      C_n^2&=0.00594(w/27)^2(10^{-5}h)^{10} \exp(-h/1000)+2.7 \nonumber \\
      &\times10^{-16}\exp(-h/1500)+A \exp(-h/100),
      \end{align}
where $A$ is a nominal value of $C_n^2(0)$ at the ground in $\text{m}^{-2/3}$, $w$ is the RMS wind spread [m/s], $h$ is the altitude, $H$ is the altitude of $S$, and $h_0$ presents the height of $R$ node above ground level.
\section{Secrecy Performance Analysis}
In this section, we derive the closed-form SOP and PPSC expressions for the proposed setup.
\subsection{Secrecy Outage Probability}
In information theoretic security, $E$ can be a malicious user which has an intention of listening the transmission. In that case, $S$ has to provide $C_s > R_s$. Thereby, SOP which is considered as an important metric in physical layer security, can be defined as the probability that the achievable secrecy capacity $C_s$ falls below a predefined threshold rate ${R_s}$, where $C_s$ can be given by 
 \begin{align}
 C_s= \begin{cases} \frac{1}{2} \log_2(1+\gamma_0)- \frac{1}{2} \log_2(1+\gamma_E)~, \gamma_0>\gamma_E \\ \\ 
0~,   \text{otherwise,}
 \end{cases}
  \end{align}
  and the SOP can be expressed as \cite{bloch2008wireless} 
\begin{align}
P_\text{SO} &= Pr\left[ C_s < R_s \right], \nonumber \\
    & = \int_{0}^{\infty}  F_{\gamma_0} \left( \gamma \gamma_{th} + \gamma_{th} - 1 \right) f_{\gamma_E} \left( \gamma \right)  d\gamma,  \\
   &\simeq \int_{0}^{\infty}  F_{\gamma_0} \left( \gamma\gamma_{th}\right) f_{\gamma_E} \left( \gamma \right)  d\gamma,\nonumber  
\end{align}
 where $\gamma_{th}=2^{2Rs}$.

In the HAPS-aided FSO-RF communication scenario, the end-to-end SNR can be written as
\small
 \begin{align}
 F_{\gamma_0} \left( \gamma \right) &= 1- \text{Pr}[\gamma_{SR}^{FSO} > \gamma] \text{Pr}[\gamma_{RD}^{RF}> \gamma] \nonumber \\
 &=1-\left(1- F_{\gamma_{SR}^{FSO}} \left( \gamma \right) \right) \left(1- F_{\gamma_{RD}^{RF}} \left( \gamma \right) \right), 
  \end{align} 
  \normalsize
and $F_{\gamma_0} \left( \gamma \right)$ can be obtained by substituting (5) and (7) into (14) as
\small
   \begin{align}
 F_{\gamma_0} \left( \gamma \right)&=1- \left( 1- B G_{r+1,3r+1}^{3r,1}\left(D \gamma\middle \vert \begin{array}{c}
1,k_1 \\
k_2,0 \end{array}
\right)\right)  \nonumber \\
&\times \sum_{k=0}^{m_{RD} -1} \sum_{i=0}^{k} \frac{\alpha_{RD}(1-m_{RD})_k.(-\delta_{RD})^k \gamma^i}{i! \lambda_{RD}^{k-i+1}\overline{\gamma}_{RD}^{k+1} .k!}  \exp (-\lambda_{RD} \gamma).
  \end{align}
\normalsize
and the PDF of $\gamma_E$ can be obtained directly from (4). Thereafter, by substituting (4) and (15) into (13), the SOP can be expressed as 
\small
\begin{align}
P_\text{SO}&=\int_{0}^{\infty} \nonumber
   \Bigg( 1-  \left( 1- B G_{r+1,3r+1}^{3r,1}\left(D \gamma_{th}\gamma\middle \vert \begin{array}{c} 1,k_1 \\
k_2,0 \end{array}\right) \right)  \nonumber  \\
 & \times \sum_{p=0}^{m_{RD} -1} \sum_{t=0}^{p} \frac{\alpha_{RD}(1-m_{RD})_p.(-\delta_{RD})^p}{t! \lambda_{RD}^{p-t+1}\overline{\gamma}_{RD}^{p+1} .p!}    \exp (-\lambda_{RD} \gamma\gamma_{th}) \Bigg)  \nonumber \\ 
&\times(\gamma \gamma_{th})^t 
\sum_{q=0}^{m_{RE} -1}  \frac{\alpha_{RE}(1-m_{RE})_q.(-\delta_{RE})^q}{ \overline{\gamma}_{RE}^{P+1} .(q!)^2}  \gamma^q \nonumber \\
&\times\exp (-\lambda_{RE} \gamma) d\gamma. 
\end{align}
\normalsize
The closed-form solution of the above-mentioned equation can be solved by using
the equation [\nocite{Wolform}21, Eq.(07.34.21.0011.01)], and with the aid of
 $\exp(-bx)=G_{0,1}^{1,0}\left(bx  \middle \vert \begin{array}{c} - \\ 0 \end{array} \right)$ as
\small
\begin{align}
P_\text{SO}&= 1-\sum_{p=0}^{m_{RD} -1} \sum_{t=0}^{p} \nonumber \frac{\alpha_{RD}(1-m_{RD})_p.(-\delta_{RD})^p}{t! \lambda_{RD}^{p-t+1}\overline{\gamma}_{RD}^{p+1} .p!}  \nonumber\\ 
&  \times\sum_{q=0}^{m_{RE} -1} \frac{\alpha_{RE}(1-m_{RE})_q.(-\delta_{RE})^q} { \overline{\gamma}_{RE}^{q+1} .(q!)^2}  \gamma_{th}^t \nonumber\\ 
 &\times\Bigg[(\lambda_{SD} \gamma_{th})^{-(q+t+1)} 
 G _{1,1}^{1,1}\left( \frac{\lambda_{SE}}{\lambda_{SD} \gamma_{th}}  \middle \vert \begin{array}{c}
-(q+t) \\ 
0 \end{array}\right)  \nonumber\\ 
&-  G_{r+2,3r+1}^{3r,2}\left(\frac{D\gamma_{th}}{(\lambda_{RE} + \lambda_{RD} \gamma_{th})}\middle \vert \begin{array}{c} 1,-(q+t),k_1 \\
k_2,0 \end{array} \right)\nonumber\\ 
&\times B (\lambda_{RE} + \lambda_{RD} \gamma_{th})^{-(q+t+1)} \Bigg].
\end{align}
\normalsize  
\vspace{0.2cm}
\subsection{Probability of Positive Secrecy Capacity Analysis}
Let us consider $E$ as a licensed user in the system which intends to capture the signal sent by the transmitter. Thus, $S$ has the information of $E$ and to guarantee information theoretic security, the condition $C_s > 0$ needs to be satisfied. Mathematically speaking, PPSC can be expressed as
\begin{align}
    P_{PPSC}&= \text{Pr} [C_s > 0] \\\nonumber 
           & = \text{Pr}\Big[\frac{1}{2} \log_2 (1+ \gamma_0) >\frac{1}{2} \log_2 (1+ \gamma_E)\Big],      \nonumber      
\end{align}
and after some manipulations, it can be given as \cite{2012probability}
\begin{align}
      P_{PPSC}=1- \int_{0}^{\infty} F_{\gamma_0} (\gamma) f_{\gamma_E}(\gamma) d\gamma.
\end{align}
Thereafter, by substituting (4) and (15) into (19) the final expression of PPSC can be obtained very similar to (17) as 
\small
\begin{align}
    P_{PPSC}&=\sum_{p=0}^{m_{RD} -1} \sum_{t=0}^{p}   \frac{\alpha_{RD}(1-m_{RD})_p.(-\delta_{RD})^p}{t! \lambda_{RD}^{p-t+1}\overline{\gamma}_{RD}^{p+1} .p!} \nonumber \\ 
& \times\sum_{q=0}^{m_{RE} -1}  \frac{\alpha_{RE}(1-m_{RE})_q.(-\delta_{RE})^q} { \overline{\gamma}_{RE}^{q+1} .(q!)^2} \gamma_{th}^t \nonumber   \\  
& \times\Bigg[(\lambda_{SD} \gamma_{th})^{-(q+t+1)} 
\times G _{1,1}^{1,1}\left( \frac{\lambda_{SE}}{\lambda_{SD} \gamma_{th}}  \middle \vert \begin{array}{c} 
-(q+t)  \\
0 \end{array}\right)  \nonumber   \\  
 &- G_{r+2,3r+1}^{3r,2}\left(\frac{D\gamma_{th}}{(\lambda_{RE} + \lambda_{RD} \gamma_{th})}\middle \vert \begin{array}{c} 1,-(q+t),k_1  \\ 
k_2,0 \end{array} \right) \nonumber\\ 
&\times B (\lambda_{RE} + \lambda_{RD} \gamma_{th})^{-(q+t+1)}  \Bigg].
\end{align}
\normalsize
\section{Numerical Results}
\small
 \begin{table}[!t]
   \renewcommand{\arraystretch}{1.1}
   \centering
\caption{List of Parameters and Values} 
\label{tab1}
\begin{tabular}{|c|c|}
 \hline
 \multicolumn{2}{|c|}{FSO link} \\
\hline  Parameter & Value  \\
\hline Optical wavelength ($\lambda$) & 1550 nm \\
\hline Satellite height ($H$) &  500 km \\
\hline HAPS altitude ($h_0$) &  14 km \\
\hline Zenith angle ($\zeta$) & 75° \\
\hline  Wind speed ($w$) & 65 m/s\\
\hline  Elevation above sea level ($h_E$) & 0.8 km \\
\hline Nominal value ($C_0$) & 1.7 $\times$ 10$^{\text{-14}}$  \\
\hline Severity of fading ($\alpha$) & 8.9033  \\
\hline Severity of scintillation ($\beta$) & 7.3955  \\
\hline Threshold ($R_s$) & 0.01 nats/s/Hz \\
\hline
\end{tabular}
\end{table}
\normalsize
\begin{figure}[!t]
  \centering
    \includegraphics[width=3.7in]{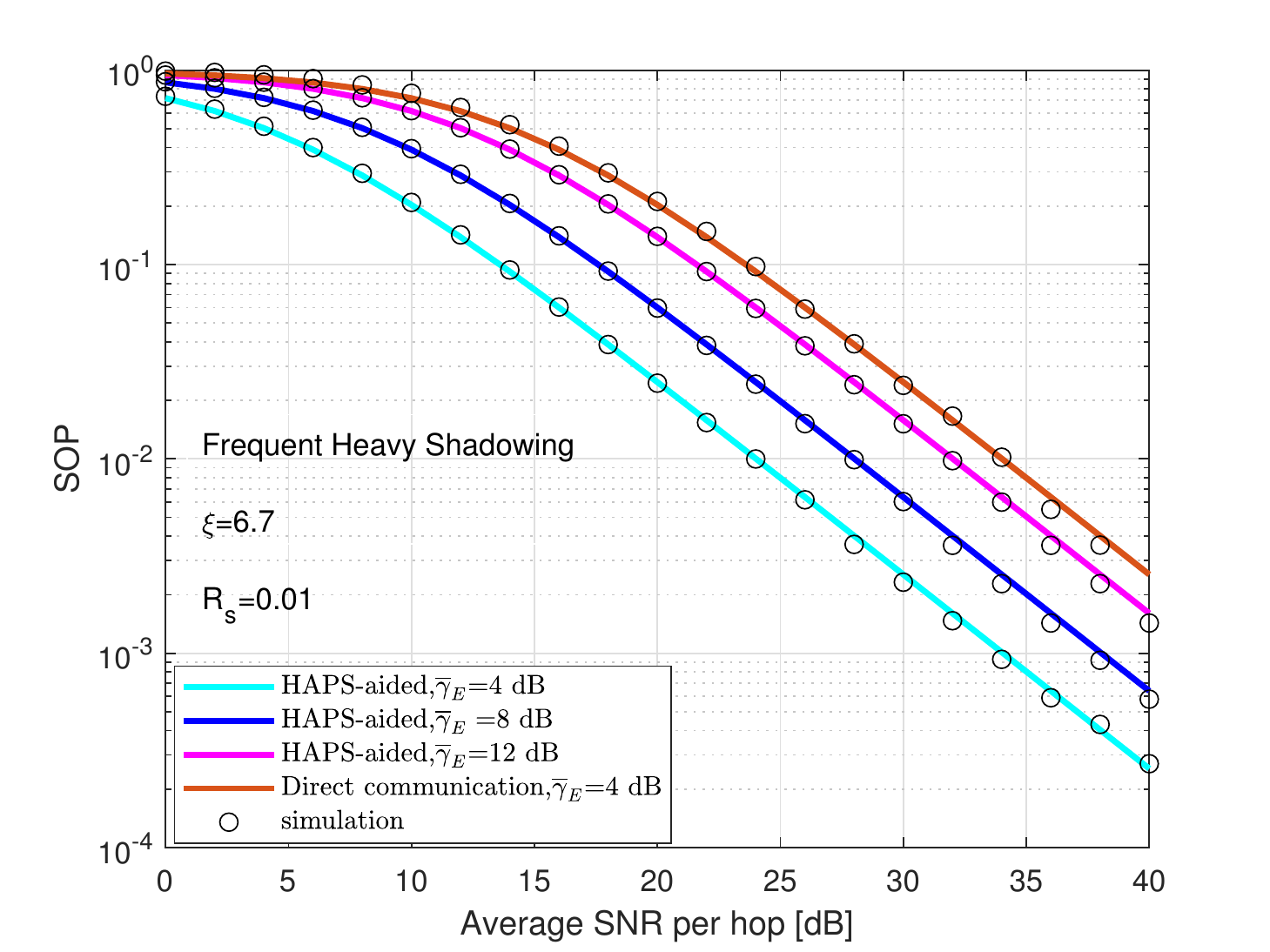}
    \caption{Secrecy outage probability performance of the proposed model for different $\overline{\gamma}_E$.}
  \label{fig:model}
\end{figure}

\begin{figure}[!t]
    \includegraphics[width=3.7in]{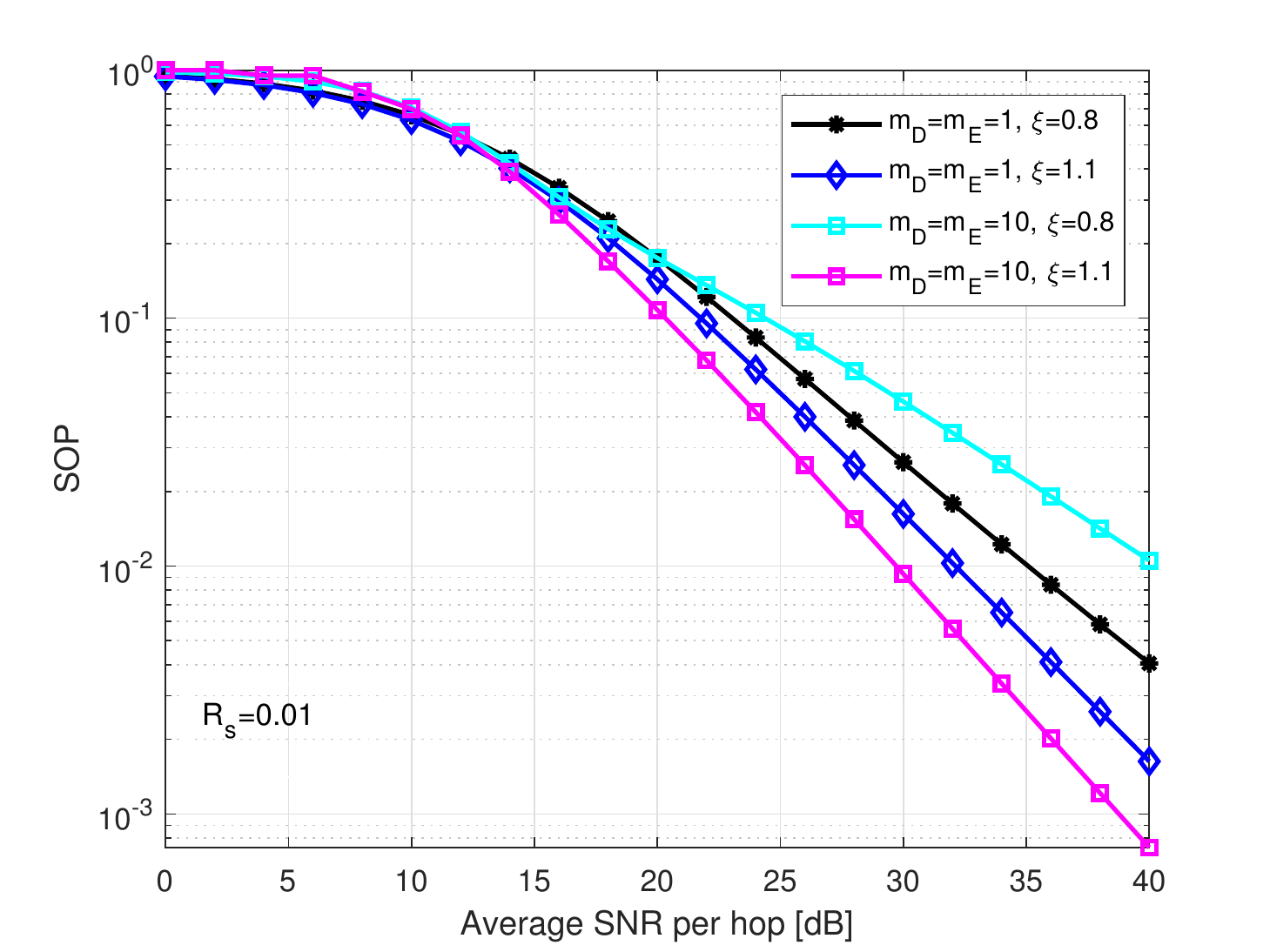}
    \caption{Impact of pointing error on the secrecy outage probability performance for different shadowing levels, $\overline{\gamma}_E=$12 dB.}
  \label{fig:model}
\end{figure}
\begin{figure}[!t]
    \includegraphics[width=3.7in]{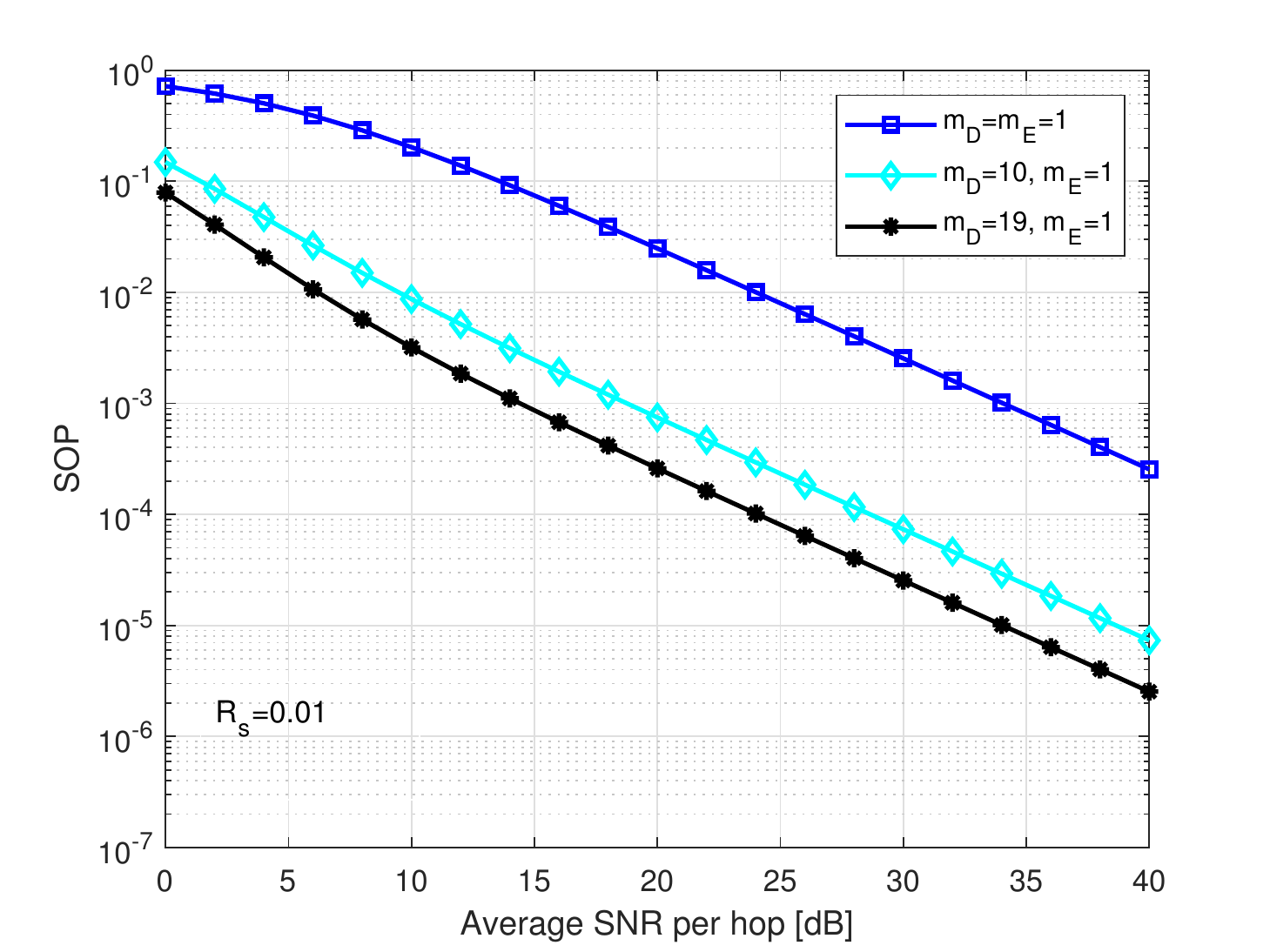}
    \caption{Secrecy outage probability performance of the proposed scheme under different shadowing levels for $D$ and $E$, $\overline{\gamma}_E=$4 dB.}
  \label{fig:model}
\end{figure}

\begin{figure}[!t]
  \centering
    \includegraphics[width=3.7in]{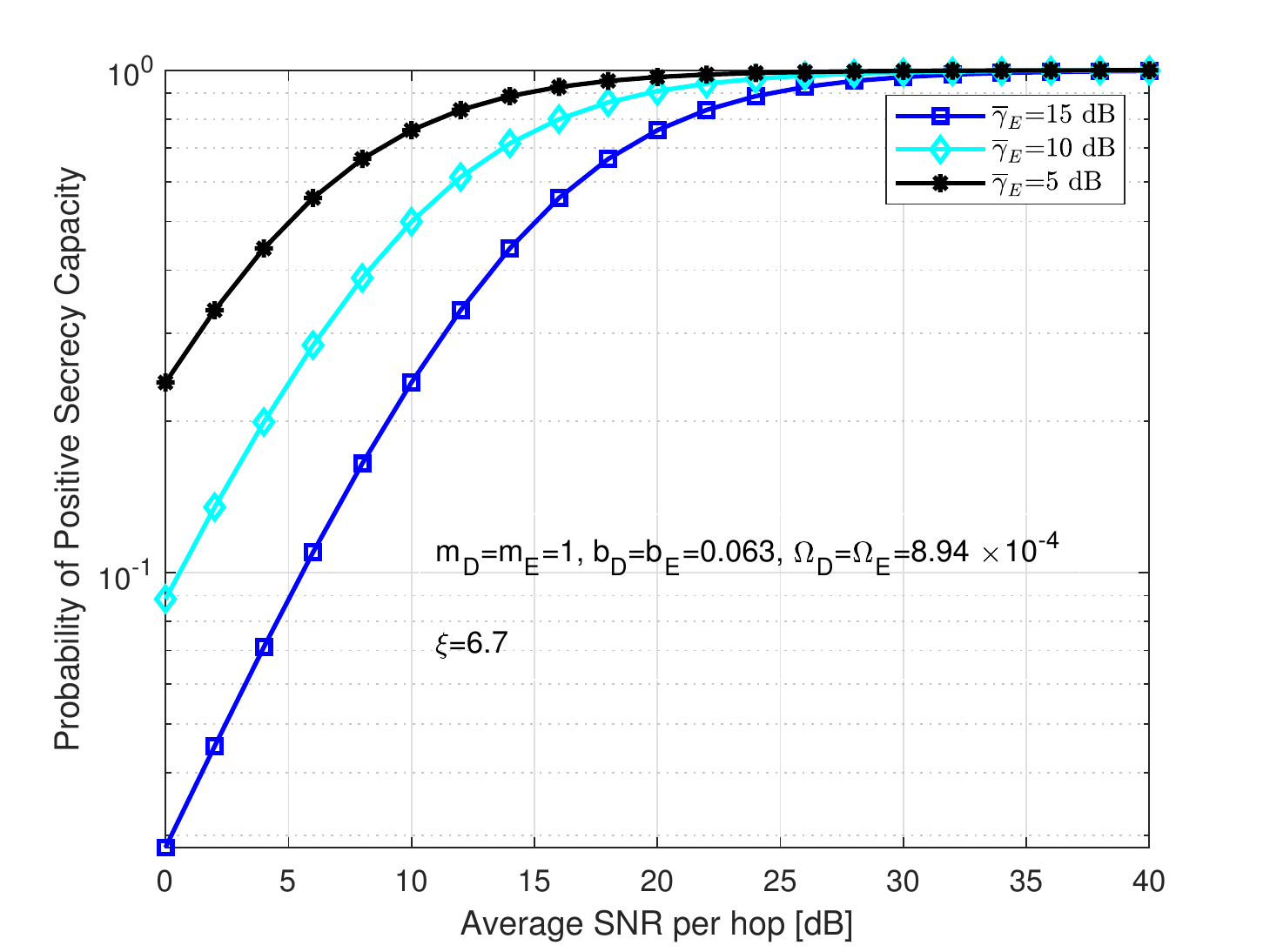}
    \caption{PPSC performance of the proposed scheme for different $\overline{\gamma}_E$ levels under frequent heavy shadowing.}
  \label{fig:model}
\end{figure}
In this section, we evaluate the secrecy performance of the proposed model under different conditions and verify our results with Monte Carlo simulations. We consider that LEO satellite is circularly orbiting at 500 km whereas, the HAPS node is at an altitude of 14 km. Furthermore, we consider different shadowing levels for the RF links: frequent heavy shadowing ($m$=1.0, $b$=0.063, $\Omega$=8.94 $\times$ 10$^{\text{-4}}$), average shadowing ($m$=10, $b$=0.126, $\Omega$=0.835), and infrequent light shadowing ($m$=19, $b$=0.158, $\Omega$=1.29) \cite{ai2019physical}. For the FSO link, we investigate the performance for HD detection ($r$=1) with $\lambda$=1550 nm, and the nominal value $A$ is $A$=1.7$\times$ 10$^{\text{-14}} \text{m}^{-2/3}$. Moreover, the wind speed level is set to $w$=65 m/s to show windy weather condition, and zenith angle is set to $\zeta$=75°. Thereby, the corresponding $\alpha$ and $\beta$ parameters are obtained as $\alpha$=8.9033 and $\beta$=7.3955.
For all figures, we consider that the threshold rate is $R_s$=0.01 nats/s/Hz. Table $\RN{1}$ presents the simulation parameters for the FSO link.
\subsection{Verification of Theoretical Expressions} 
As we can see, Fig. 2 illustrates the SOP as a function of the average SNR per-hop $\overline{\gamma}$ under different eavesdropper SNRs for frequent heavy shadowing. The theoretical curves, which are illustrated with solid lines are verified with the simulations which are shown with circles. Furthermore, we can observe the good agreement with the simulations which confirms our derivations. It is obvious from this figure that larger $\overline{\gamma}_E$ values imply lower SOP performance. Furthermore, it is clear from the figure that the proposed scenario outperforms the scenario of using a direct RF link due to path loss caused by the large distance between the satellite and the ground station. 

In Fig. 3, we investigate the pointing error effect for frequent heavy shadowing and average shadowing. As we can see from the figure, in the case of $\xi$=1.1 that indicates a lower level of pointing, the SOP under average shadowing outperforms the SOP under frequent heavy shadowing. However, when $\xi$=0.8 the overall performance of the SOP under average shadowing deteriorates. In fact, as the link between the satellite and HAPS is highly affected due to the misalignment caused by a high level of pointing error, it dominates the communication even though the channel between HAPS to ground is improved.

In Fig. 4, we evaluate the SOP when the main destination is under different shadowing levels, whereas the eavesdropper is under frequent heavy shadowing. As expected, when the channel of the intended receiver is less shadowed than the channel of the illegitimate receiver, the overall performance of the SOP enhances.

Fig. 5 shows the PPSC performance of the RF eavesdropper attack for different $\gamma_E$ values when both channels of $D$ and $E$ are under frequent heavy shadowing, for $\xi$=6.7. As observed from the figure, the PPSC increases as $\gamma_E$ decreases. This shows the high impact of $\gamma_E$ to guarantee secure communication.
\subsection{Design Guidelines}
Finally, we outline some important points that can be used in the design of secure downlink SatCom.
\begin{itemize}
\item For the design of the RF link, severe shadowing conditions can deteriorate the overall secrecy performance of HAPS-aided FSO-RF SatCom.

\item Considering that the HAPS node is located at the stratosphere, the FSO channel from the satellite to the HAPS can not be affected heavily from the adverse weather conditions. Thereby, HAPS can guarantee the secure communication even in the presence of harsh winds and the overall SOP performance can be enhanced.

\item The simulations have shown that higher values of pointing error enhance the overall performance. Thus, $\xi$ can be considered as an important parameter in the design of the FSO communication as it is related with the misalignment of the communicating parts.

\item The average SNR of the eavesdropper can be considered as a crucial parameter in the scenario to secure the HAPS-aided SatCom.
\end{itemize}
\vspace{0.2cm}
\section{Conclusion}
In this paper, a framework to counter RF eavesdropping has been proposed by considering a HAPS aided FSO-RF SatCom. Specifically, we derived the closed-form of secrecy outage probability and the probability of positive secrecy capacity expressions for the proposed system model. The results, which were validated with simulations, proved that the RF link conditions have a larger impact on the performance of the investigated system. Furthermore, shadowing parameters, and the pointing errors can be critical to guarantee the secure SatCom. As expected, when the channel of the illegitimate user is more shadowed, the secrecy performance is enhanced. Finally, we provided some important design guidelines that can be helpful for the HAPS-aided downlink SatCom systems.
\vspace{0.3cm}
 \balance
\bibliographystyle{IEEEtran}
\vspace{0.3cm}
\bibliography{main}
\end{document}